\documentclass[conference]{IEEEtran}
\IEEEoverridecommandlockouts
\usepackage{cite}
\usepackage{amsmath,amssymb,amsfonts}
\usepackage{algorithmic}
\usepackage{graphicx}
\usepackage{textcomp}
\usepackage{xcolor}
\usepackage{hyperref}
\def\BibTeX{{\rm B\kern-.05em{\sc i\kern-.025em b}\kern-.08em
    T\kern-.1667em\lower.7ex\hbox{E}\kern-.125emX}}
\begin{document}

\title{An Intelligent Mechanism for Monitoring and Detecting Intrusions in IoT Devices}

\author{\IEEEauthorblockN{Vitalina Holubenko}
\IEEEauthorblockA{\textit{CISUC, DEI, University of Coimbra} \\
Coimbra, Portugal \\
vitalina@student.dei.uc.pt}
\and
\IEEEauthorblockN{Paulo Silva}
\IEEEauthorblockA{\textit{LIS, Instituto Pedro Nunes} \\
Coimbra, Portugal \\
pmgsilva@ipn.pt}
\and
\IEEEauthorblockN{Carlos Bento}
\IEEEauthorblockA{\textit{CISUC, DEI, University of Coimbra} \\
Coimbra, Portugal \\
bento@dei.uc.pt}
\and
}


\maketitle

\begin{abstract}

The current amount of IoT devices and their limitations has come to serve as a motivation for malicious entities to take advantage of such devices and use them for their own gain. To protect against cyberattacks in IoT devices, Machine Learning techniques can be applied to Intrusion Detection Systems. Moreover, privacy related issues associated with centralized approaches can be mitigated through Federated Learning. This work proposes a Host-based Intrusion Detection Systems that leverages Federated Learning and Multi-Layer Perceptron neural networks to detected cyberattacks on IoT devices with high accuracy and enhancing data privacy protection.

\end{abstract}

\begin{IEEEkeywords}
Intrusion Detection System, Federated AI, Machine Learning, Internet of Things, Security, Privacy
\end{IEEEkeywords}

\section{Introduction and Background}

Intrusion Detection Systems (IDS) are an indispensable form of defence mechanisms that examine activities within a system or a network, to identify and alert about incoming attacks. IDS have been studied extensively, however, little work has been done with HIDS for IoT, which we aim to tackle in this work. Traditional IDS are not suitable for IoT for many reasons, such as the limited resources on such devices and their decentralization. In addition, there is considerable heterogeneity in the devices, technologies and network protocols used in IoT. In HIDS, the data from the host system’s audit and logging mechanisms are analysed to look for signs of intrusions. System call traces in particular are often used in detecting intrusions with HIDS. System call traces are used to find behavioural patterns, enabling intrusion detection during execution. A system call trace refers to the system call sequences that have been ordered, performed by a process that a program (i.e., process) ran during execution. The system call analysis approach to intrusion detection was first proposed by Forrest et al. \cite{b10}, where short sequences of system calls are used to generate profiles of normal program behaviour. Other works \cite{b11}\cite{b12} tried to improve on the results obtained by Forrest by using machine learning algorithms to extract information from normal and abnormal sequences of system call traces.

In this work we leverage Federated Learning (FL) for the proposed HIDS. It is common, in traditional Machine Learning (ML) scenarios, to have a centralized server, where the data from several devices (i.e., smartphones, tablets, laptops) is aggregated. Such devices collect vast amounts of data, which inherently represents a privacy risk. FL (introduced by Google in 2016) is suitable for building distributed IoT systems due to the recent advances in mobile technologies and the overgrowing concerns of risks related to user privacy. In this case, FL enables the cooperative training of a shared global model, which benefits both network operators and IoT clients in terms of network resource savings and privacy enhancements because user data is not shared to a central entity.

As IoT-oriented systems are more vulnerable targets of cyberattacks \cite{b9}, we have studied host intrusion detection methods that are compatible with IoT devices and HIDS approach. We have performed a state-of-the-art analysis of IDS, their characteristics, typical architectures, and the types of data that could be used for intrusion detection. Ultimately, we propose an AI-based Host Intrusion Detection System (HIDS) for IoT devices, the main objective of this work. 

\section{Experimental Design}

The main focus of this work is to propose a HIDS approach for IoT devices. The proposed system should enable deployment in lightweight IoT devices with significantly less computation power than classical devices or computer systems. To be deployed, the solution needs to required the minimum possible storage and run on restrained memory resources.

\begin{figure}[!t]
    \centering
    \includegraphics[width=1\columnwidth]{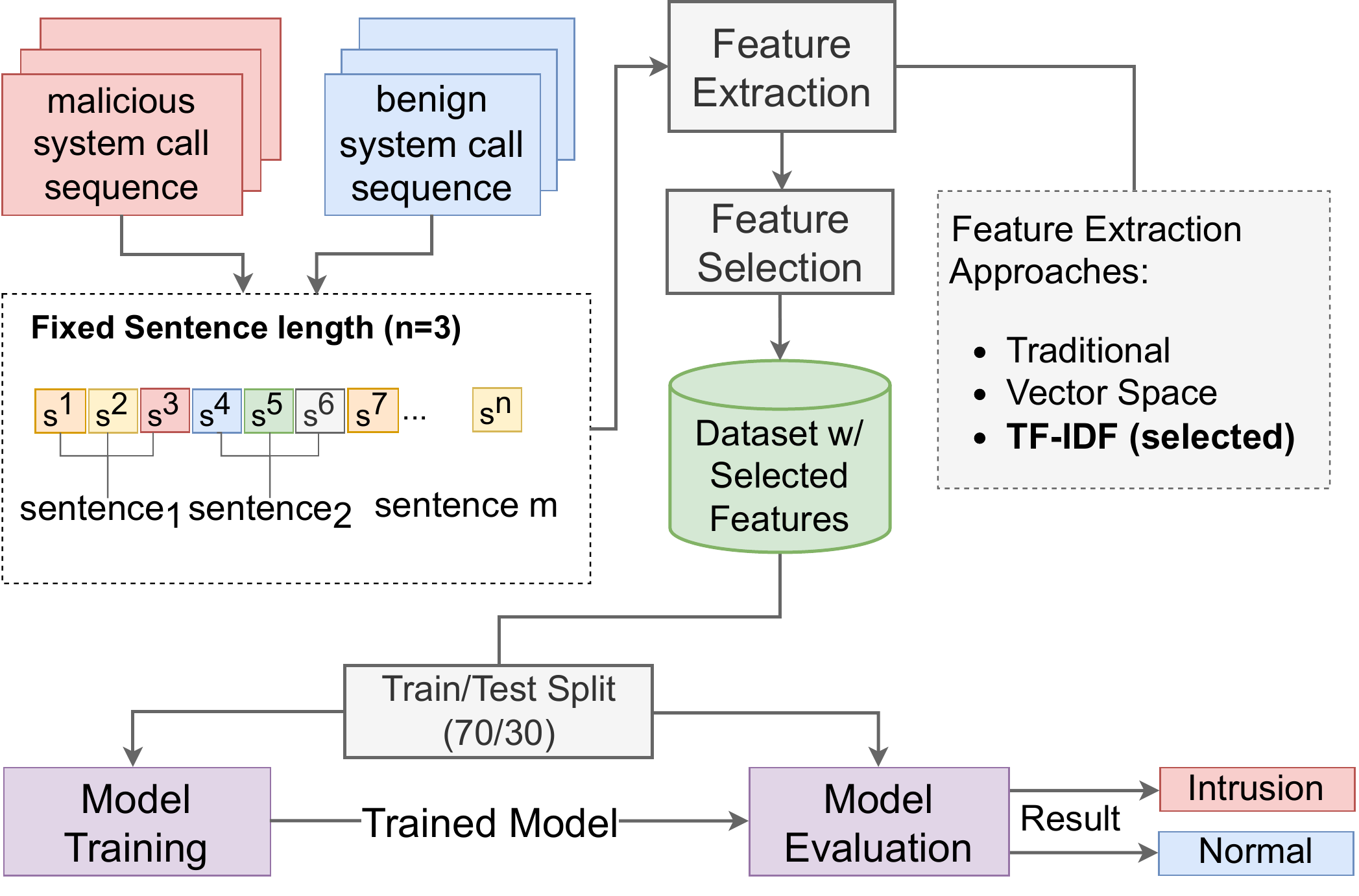}
    \caption{Proposed Approach}
    \label{fig:architectture}
\end{figure}

The proposed architecture for the ML HIDS is represented in Figure \ref{fig:architectture}. It starts with the data extraction module, where sequences are recorded in the database as part of the normal profile, these profiles usually consist of thousands of short sequences of system calls. The data extraction module aggregates the system call traces, which are going to be processed via feature extraction ((i) trivial, (ii) vector space and (iii) TF-IDF representation) and feature selection (via Principal Component Analysis).

In the generation of anomalous data phase, we simulate an environment of the OS which is under intrusion and capture its behaviour under those circumstances, which is used in order to train a model that recognizes anomalous device behaviour. At different stages of this work we have also a publicly available dataset (ADFA-LD).

The processed data is then subject of automated analysis by a ML model  that classifies the input as anomalies or normal, and then raises an alert when an intrusion is detected. 

Two training ML approaches were used for evaluation: centralized model and a decentralized FL-based approach, where each node represents an IoT client. The main objective of this approach is to establish a comparison between the traditional centralized ML approaches and the federated setting.

In the evaluation of ML models, we employed two datasets. The first (ADFA-LD) is a collection of system call traces with both benign and malicious syscall traces - a single trace is a sequence of system call numbers, over some arbitrary time period. The second, is a proof of concept version of an HIDS dataset based on system calls, with normal traces and attack traces (on local VMs).

In the training phase, the class were balanced as 50\% benign traffic and 50\% attack traffic, making the dataset balanced for binary classification. Although it is important to note that this balance of the data can lead to different problems, since in real world scenarios the amount of benign data surpasses significantly malicious data. Both train and test sets are indeed affected by each change, and the goal is to compare the impact on the model performance when the classes are balanced. 

\section{Experimental Results}

\begin{figure}[!t]
    \centering
    \includegraphics[width=1\columnwidth]{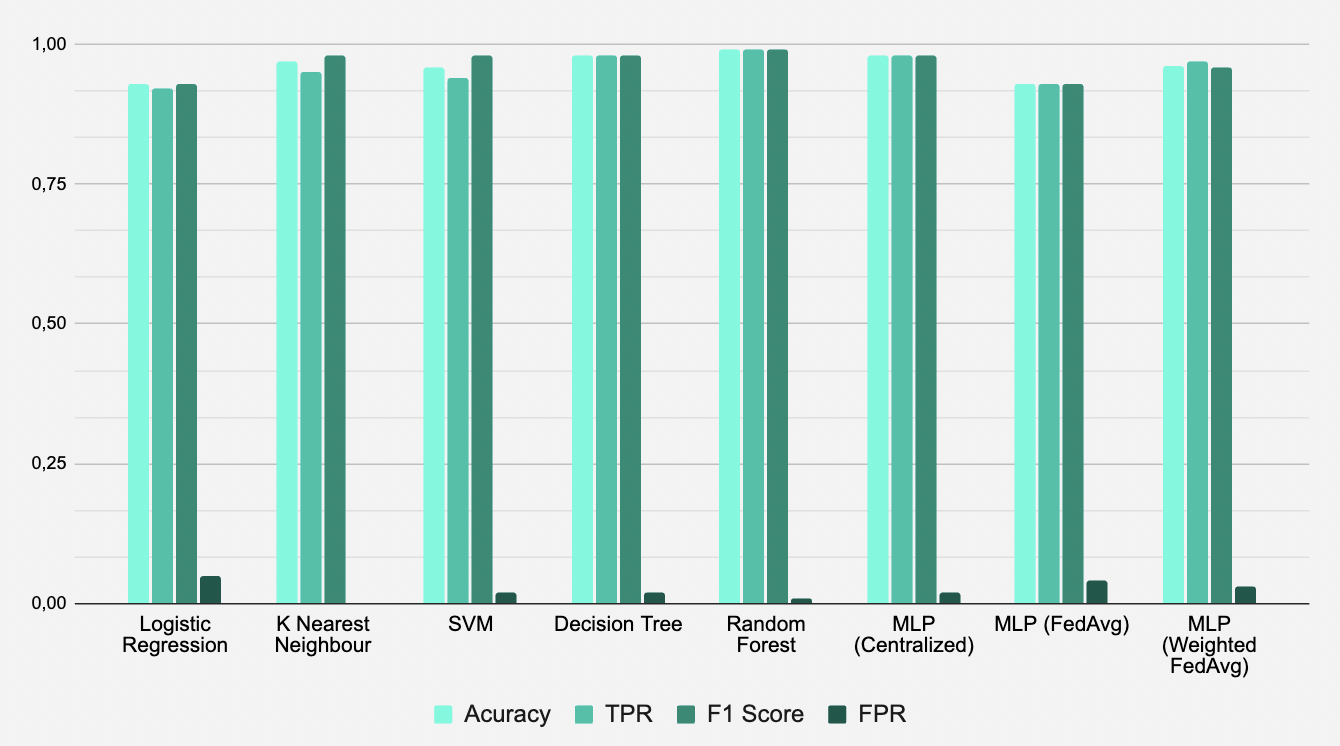}
    \caption{Accuracy, TPR, F1 Score and FPR values with TF-IDF}
    \label{fig:results}
\end{figure}

Figure \ref{fig:results} shows the performance of different machine learning algorithms on ADFA-LD dataset, with TF-IDF representation, as it has shown to be better suited for model training. Our goal is not only to find the best candidate model with high accuracy (high recall but low FPR and FNR), but also the most suitable length for the system call traces, which after experimenting with various sequence lengths, we concluded that 30 was the optimal length. Decision Tree, Random Forest, KNN and MLP are the best candidate algorithms since they achieved higher accuracy, recall and precision, yet at a lower false positive rate. In terms of FPR and FNR, the results reveal that Random Forest, SVM, KNN, and MLP models outperform other classifiers by a considerable margin, up to 0.05\%. Random Forest was the best performing classifier, obtaining a value of 98\% for F1-score.

In regard to the FL setting, the performance of the algorithm has shown to be promising when comparing to the baseline, achieving around 96\% of accuracy. The performance of the two federated algorithms was similar, however, when the data is not equally distributed, we can see that Weighted Federated Averaging (WFA) has a slightly better performance than Federated Averaging (FA), and is able to reach better accuracy with the scaling of the weights.

\section{Conclusion}


The proposed IDS architecture, evaluated in terms of accuracy, precision and recall, proved to be effective in classifying system calls into benign and real time malicious samples with various machine learning algorithms, in particular MLP-based neural networks. The current working system is able to achieve a F1-score of 96\%. Additionally, with Federated Learning, the accuracy of the system can be improved overtime trough local training of the models.

The developed HIDS can be further enhanced by implementing other efficient machine learning algorithms, such as deep learning techniques, like CNN, RNN and LSTM as well as multi class classification in order to differentiate the incoming attacks. 

\section*{Acknowledgment}

This work was carried out in the scope of the ARCADIAN-IoT - Autonomous Trust, Security and Privacy Management Framework for IoT, Grant Agreement Number: 101020259. H2020-SU-DS02-2020.

\end{document}